\documentclass[twocolumn]{fel99}
\usepackage{latexsym}
\begin{document}
\pagestyle{empty}
\begin{frontmatter}
\noindent
\vskip -14mm

\title {\bf EXCITATION OF RESONATORS BY ELECTRON BEAMS}
\author {E.G.Bessonov$^{\dagger \dagger }$\thanksref{1}}
\author{Yukio Shibata$^{\dagger}$, Satoshi Sasaki$^{\dagger}$,
Kimihiro Ishi$^{\dagger}$, Mikihiko Ikezawa$^{\dagger}$}
\vskip -2mm
\address {$^{\dagger}$ RISM, Tohoku University, Japan
$^{\dagger \dagger }$ Lebedev Phys. Inst. RAS, Moscow, Russia}
\thanks[1]{Corresponding author. bessonov@sgi.lpi.msk.su.}
\vspace{-7mm}
                     \begin{abstract}
Elements of a little-known vector theory of open resonators and
experiments on excitation of a fundamental mode with transverse and
longitudinal polarization in such resonators are discussed.

\vspace{-1mm}
\noindent{\footnotesize \it PACS number(s): 41.60.Cr.}
\hskip 5mm
\noindent{\footnotesize Keywords: Open resonators, Longitudinal fundamental
TEM mode, polarization.} \end{abstract}

\end{frontmatter}
\vskip -7mm
\small

The propagation of electromagnetic waves in vacuum is described by
Maxwell equations. More simple wave equations for electromagnetic
fields coupled by conditions $div \vec E = div \vec H =0$, or for the
potentials $\vec A$, $\varphi$ introduced by the equations $\vec H =
rot \vec A$, $\vec E = - grad \varphi - (1/c)(\partial \vec A/\partial
t)$ and coupled by a gauge condition can be used for a simplified
solution of electrodynamic problems. The following simplification is
using the electric and magnetic Hertz vectors $\vec \Pi ^e$, $\vec
\Pi ^m$.  They are introduced by $\vec{A}= (1/c)(\partial
\vec \Pi ^{e/m})/(\partial t), \hskip 3mm \varphi = - div \vec \Pi
^{e/m}$.  The so defined potentials $\vec A$ and $\varphi$ will
satisfy both the gauge and Maxwell equations. In this case, the
electric and magnetic field strengths are

\vskip -5mm

        $$\vec{E} = grad\, div \Pi ^e - {1\over c ^2}\frac
        {\partial ^2 \vec \Pi ^e}{\partial t^2} - {1\over
        c}{\partial\over \partial t} rot \vec \Pi ^m,$$
        \begin{equation} 
        \vec H = {1\over c} {\partial \over \partial t} rot \vec \Pi ^e
        + grad\, div \vec \Pi ^m - {1\over c ^2}\frac {\partial ^2 \vec
        \Pi ^m}{\partial t^2}.  \end{equation}

The vectors $\vec \Pi ^e$ and $\vec \Pi ^m$ fulfil the wave
equation $\Box \vec \Pi ^{e/m} = 0$. We can identify the solution of
the scalar wave equation $\Box U = 0$ with one component of vectors
$\vec \Pi ^e$ or $\vec \Pi ^m$ (e.g. $\vec \Pi ^e = \vec e _x \cdot 0 +
\vec e _y \cdot 0 + \vec e _z \cdot U,$\hskip 2mm $ \vec \Pi ^m = 0$).
Substituting the vector in (1) we find the electromagnetic
field strengths. Then we can identify the same solution $U$ by
another component of the Hertz vector, equate the rest components to
zero, and calculate the other electromagnetic field strengths. After
going through all the compositions of components we obtain a
set of six different electromagnetic wave modes \cite{oraevskiy}.

The monochromatic light beams of a limited diameter related to the
resonator modes can be written in the form $U = V(x,y,z)e^{i(kz -\omega
t)}$, where $V(x,y,z)$ is the function of a coordinate slowly varying
in comparison with $\exp[i(kz - \omega t)]$. In a paraxial
approximation $|\partial^2 V/\partial z ^2| \ll 2k|\partial V/\partial
z|$, $k = \omega /c$ this form is described by the equation $2k
i{\partial V/ \partial z} + {\partial ^2 V/ \partial x^2} + {\partial
^2 V/ \partial y^2} = 0$.  The solution of this equation by the method
of separation of variables in the cylindrical coordinates, where
$V(x,y,z) = G(u)\Phi (\phi)\exp [ikr ^2/2q(z)]\cdot \exp[iS(z)]$, $r =
\sqrt {x ^2 + y ^2}$, and $\phi$ are the cylindrical coordinates, $u =
r/w(z)$, has the form

\vskip -3mm
        $$V(r,\phi,z) = $$ $${C \over w(z)} \left({r \over
         w(z)}\right) ^m{\sin m\phi \choose \cos m\phi}L ^m _n\left({2r
        ^2\over w ^2(z)}\right)$$

        \begin{equation} 
        \exp \left\{{ikr^2\over 2q(z)} - i(m + 2n +1)
        arctg {z\over Z _R}\right\},
        \end{equation}
where $L ^m_n$ are the Lagerian polynomials ($L ^0 _0 (\xi) = 1, L ^0
_1 = 1 - \xi,$ ...); $Z _R = \pi w _0 ^2/ \lambda$, the Rayleigh length;
$\lambda = 2\pi c/\omega$, the wavelength; $C$ = constant; ${1/ q(z)} =
{1/ R(z)} + {i\lambda / \pi w ^2}; R(z) = z[1 + ({Z _R/\, z}) ^2]$, the
radius of the wave front of Gaussian beam; $w ^2(z) = w _0^2[1 + ({z/ Z
_R}) ^2]$; $w (z)$, the radius of the beam, and $w _0(z)$ the radius of
the beam waist.

The compositions 1) $\Pi ^e_x = U(x,y,z)$, $\Pi ^e_y = \Pi ^e_z = 0$ or
2) $\Pi ^e_x = 0$, $\Pi ^e_y = U(x,y,z)$, $\Pi ^e_z = 0$ or 3) $\Pi
^e_x = 0$, $\Pi ^e_y = 0$, $\Pi ^e_z = U(x,y,z)$ together with the
conditions $\vec \Pi ^m = 0$, $\partial ^2V/\partial x_i \partial x_k
\ll k\partial V/\partial x_i \ll k^2V$ lead to the field strengths

\noindent
\vspace{ 1mm}
$$E^{e,1}_x \simeq - H^{e,1} _y \simeq k^2U(x,y,z), \hskip 5mm
E^{e,1} _y \simeq H ^{e,1}_x \simeq 0, $$ $$ E ^{e,1} _z =
        2ikx\left[{1\over w^2(z)} + {ik\over R(z)} \right] U(x,y,z),$$
        $$H_z^{e,1} = 2iky \left[{1\over w^2(z)} + {ik\over R(z)}
        \right] U(x,y,z)$$ $$ E^{e,3}_x = - H ^{e,3} _y =
        2ikx\left[{1\over w^2(z)} + {ik\over R(z)}\right]U(x,y,z),$$
        $$ E ^{e,3} _y = - H ^{e,3} _x = - 2iky\left[{1\over w ^2 (z)}
        + {ik\over R(z)}\right]U(x,y,z),$$
        $$E ^{e,3} _z = 2ik[{w _0 ^2
        z\over w^2(z) Z _R ^2}({2r ^2 \over w ^2 (z)} - 1) -
        {ikr^2\over 2R^2(z)} $$
         \begin{equation} 
        (1 - {Z _R ^2\over z ^2}) - {i w _0
        ^2 \over w ^2 Z _R }]U(x,y,z),\hskip 5mm H_z^3 = 0,
         \end{equation}
where the upper superscripts show the composition of the electric Hertz
vector corresponding to the transverse (x,y) and longitudinal (z)
polarizations. The second case can be received from the first one by
substitution of the variable $x$ by $y$, and vice versa. The field
strengths received from the magnetic Hertz vector are  $\vec E ^{'} = -
\vec H$, $\vec H ^{'} = \vec E$.

In the first case, the Gaussian beam has mainly the transverse field 
components, where $E ^{e,1} _z (r = 0) = 0$. In the third case, the 
field components are compatible by the value ($E ^{e,1} _z (r = 0) \ne 
0$) \cite {oraevskiy}, \cite {davis}. These are the fundamental 
transversely and longitudinally polarized $TEM ^{e,1} _{00}$ and $TEM 
^{e,3} _{00}$ ($TM _{01}$) modes, accordingly. The $TEM ^{e,3} _{00}$ 
modes can be excited by the transition radiation emitted on mirrors of 
open resonators by electrons that are homogeneously moving along their 
axes.  Such an excitation was probably observed in the experiments 
published in \cite{shibata}. Previous experiments on excitation of open 
resonators were done under conditions when the electron trajectories 
were directed either at some angle to the axis of a resonator \cite 
{brannen}, or along caustics of the fundamental mode of a resonator 
\cite {chernenko}, where the fundamental $TEM _{00} ^{e,1}$ modes had  
not a longitudinal component of the electric field strength at the open 
resonator axis. Electron beams of finite transverse dimensions moving 
along the exes of open resonators can excite non-efficiently higher 
modes like $TEM _{10} ^{e,1,2}$ having small longitudinal components 
near the exes of resonators \cite [b]{shibata}.

Excitation of the $TEM ^{e,3} _{00}$ modes is possible at even
harmonics of undulator radiation in free-electron lasers using the flat
undulators with a high deflecting parameters when the amplitudes of
longitudinal oscillations are high. Radiation stored at this mode
in supercavity can be used for laser driven acceleration in vacuum as
well.

We thank Prof. A.N.Oraevsky and Prof. Ming Xie  for the discussion of
this paper.

\end{document}